\documentstyle[]{mn}
\newif\ifAMStwofonts

\def\etal{{\rm et al.}}

\def\simgt{\mathrel{\spose{\lower 3pt\hbox{$\sim$}}
        \raise 2.0pt\hbox{$>$}}}
\def\simlt{\mathrel{\spose{\lower 3pt\hbox{$\sim$}}
        \raise 2.0pt\hbox{$<$}}}

\ifoldfss
  \newcommand{\rmn}[1] {{\rm #1}}

  \ifCUPmtlplainloaded \else
    \NewTextAlphabet{textbfit} {cmbxti10} {}
    \NewTextAlphabet{textbfss} {cmssbx10} {}
    \NewMathAlphabet{mathbfit} {cmbxti10} {} 
    \NewMathAlphabet{mathbfss} {cmssbx10} {} 
  \fi
  \ifAMStwofonts
    \ifCUPmtlplainloaded \else
      \NewSymbolFont{upmath} {eurm10}
      \NewSymbolFont{AMSa} {msam10}
      \NewMathSymbol{\upi}     {0}{upmath}{19}
      \NewMathSymbol{\umu}     {0}{upmath}{16}
      \NewMathSymbol{\upartial}{0}{upmath}{40}
      \NewMathSymbol{\leqslant}{3}{AMSa}{36}
      \NewMathSymbol{\geqslant}{3}{AMSa}{3E}

    \fi
  \fi
\fi 

\ifnfssone
  \newmathalphabet{\mathit}
  \addtoversion{normal}{\mathit}{cmr}{m}{it}
  \addtoversion{bold}{\mathit}{cmr}{bx}{it}
  \newcommand{\rmn}[1] {\mathrm{#1}}

  \newmathalphabet{\mathbfit} 
  \addtoversion{normal}{\mathbfit}{cmr}{bx}{it}
  \addtoversion{bold}{\mathbfit}{cmr}{bx}{it}
  \newmathalphabet{\mathbfss} 
  \addtoversion{normal}{\mathbfss}{cmss}{bx}{n}
  \addtoversion{bold}{\mathbfss}{cmss}{bx}{n}
  \ifAMStwofonts
    \ifCUPmtlplainloaded \else
      \UseAMStwoboldmath
      \makeatletter
      \new@mathgroup\upmath@group
      \define@mathgroup\mv@normal\upmath@group{eur}{m}{n}
      \define@mathgroup\mv@bold\upmath@group{eur}{b}{n}
      \edef\UPM{\hexnumber\upmath@group}
      \new@mathgroup\amsa@group
      \define@mathgroup\mv@normal\amsa@group{msa}{m}{n}
      \define@mathgroup\mv@bold\amsa@group{msa}{m}{n}
      \edef\AMSa{\hexnumber\amsa@group}
      \makeatother
      \mathchardef\upi="0\UPM19
      \mathchardef\umu="0\UPM16
      \mathchardef\upartial="0\UPM40
      \mathchardef\leqslant="3\AMSa36
      \mathchardef\geqslant="3\AMSa3E
    \fi
  \fi
\fi 

\ifnfsstwo
  \newcommand{\rmn}[1] {\mathrm{#1}}

  \DeclareMathAlphabet{\mathbfit}{OT1}{cmr}{bx}{it}
  \SetMathAlphabet\mathbfit{bold}{OT1}{cmr}{bx}{it}
  \DeclareMathAlphabet{\mathbfss}{OT1}{cmss}{bx}{n}
  \SetMathAlphabet\mathbfss{bold}{OT1}{cmss}{bx}{n}
  \ifAMStwofonts
    \ifCUPmtlplainloaded \else
      \DeclareSymbolFont{UPM}{U}{eur}{m}{n}
      \SetSymbolFont{UPM}{bold}{U}{eur}{b}{n}
      \DeclareSymbolFont{AMSa}{U}{msa}{m}{n}
      \DeclareMathSymbol{\upi}{0}{UPM}{"19}
      \DeclareMathSymbol{\umu}{0}{UPM}{"16}
      \DeclareMathSymbol{\upartial}{0}{UPM}{"40}
      \DeclareMathSymbol{\leqslant}{3}{AMSa}{"36}
      \DeclareMathSymbol{\geqslant}{3}{AMSa}{"3E}
    \fi
  \fi
\fi 

\ifCUPmtlplainloaded \else
  \ifAMStwofonts \else 
    \def\upi{\pi}
    \def\umu{\mu}
    \def\upartial{\partial}
  \fi
\fi

\title[Interpretation of the OGLE Q2237+0305 microlensing light-curve]
  {Interpretation of the OGLE Q2237+0305 microlensing light-curve (1997-1999)}
\author[J. S. B. Wyithe et al.]
  {J.~S.~B.~Wyithe$^{1,2}$, 
  R.~L.~Webster$^1$, 
  E.~L.~Turner$^2$ \\
  $^1$ School of Physics, The University of Melbourne, Parkville, Vic, 3052, Australia\\
  $^2$ Princeton University Observatory, Peyton Hall, Princeton, NJ 08544, USA\\
 Email: swyithe@astro.Princeton.edu, rwebster@physics.unimelb.edu.au, elt@astro.Princeton.edu }
\date{Accepted. Received}
\pagerange{\pageref{firstpage}--\pageref{lastpage}}
\pubyear{1999}

\def\LaTeX{L\kern-.36em\raise.3ex\hbox{a}\kern-.15em
    T\kern-.1667em\lower.7ex\hbox{E}\kern-.125emX}

\begin{document}

\label{firstpage}

\maketitle

\begin{abstract}
The four bright images of the gravitationally lensed quasar Q2237+0305 are being monitored from the ground (eg. OGLE collaboration, Apache Point Observatory) in the hope of observing a high magnification event (HME). Over the past three seasons (1997-1999) the OGLE collaboration has produced microlensing light-curves with unprecedented coverage. These demonstrate smooth, independent (therefore microlensing) variability between the images (Wozniak et al. 2000a,b; OGLE web page). We have retrospectively applied probability functions for high-magnification event parameters to the observed light-curve features. We conclude that the 1999 image C peak was due to the source having passed outside of a cusp rather than to a caustic crossing. In addition, we find that the image C light-curve shows evidence for a caustic crossing between the 1997 and 1998 observing seasons involving the appearance of new critical images. Our models predict that the next image C event is most likely to arrive $500$ days following the 1999 peak, but with a large uncertainty ($\sim 100-2000$ days). Finally, given the image A light-curve derivative at the end of the 1999 observing season, our modelling suggests that a caustic crossing will occur between the 1999 and 2000 observing seasons, suggesting a minimum for the image A light-curve $\sim 1-1.5$ magnitudes fainter than the November 1999 level.
\end{abstract}


\section{Introduction}

The QSO 2237+0305, sometimes known as Huchra's lens or the Einstein Cross (Huchra et al. 1985), is perhaps the most remarkable gravitational lens yet discovered. It comprises a foreground barred Sb galaxy (z=0.0394) whose nucleus is surrounded by four images of a radio-faint QSO (z=1.695). Ground based spectroscopic observations have verified that all four images are similar QSO's at the same redshift (Adam et al. 1989). Broad band monitoring has shown that significant microlensing events occur (eg. Irwin et al. 1989; Corrigan et al. 1991). Since the optical depth to microlensing is of order unity at each of the image positions (eg. Kent \& Falco 1988; Schneider et al. 1988; Schmidt, Webster \& Lewis 1998), the magnification effects on the source can be considered as a network of caustics moving across the source plane. Strong variation in a particular image results from the source either crossing a caustic or passing close to a cusp. Q2237+0305 provides a unique opportunity to study microlensing events for two good reasons. Firstly, the relative closeness of the lensing galaxy and the nearly on-axis alignment means that the time delay between the four quasar images is less than a day. Thus it is easy to separate microlensing events from intrinsic variations of the QSO. Also, for this QSO the time-scale for microlensing events is typically 30-50 days, reduced by a factor of 10 from typical time-scales of perhaps a year.

Although Q2237+0305 has been monitored since its discovery, data collected by the OGLE collaboration over the last 3 years (Wozniak et al. (2000a,b); see also http://www.astro.princeton.edu/$\sim$ogle/ogle2/huchra.html) has for the first time obtained data of sufficient coverage to clearly demonstrate smooth independent flux variation between the images. In this paper we provide interpretations of several features in the image A and C light-curves, and discuss their implications for future microlensing. Secs. \ref{data} and \ref{model} describe the published light-curves and the models used to interpret them. In Sec. \ref{OGLE_lc_sec} we discuss the HME classes associated with different light-curve features by comparing their parameters (such as height and maximum light-curve derivative) to model distributions. We then describe the implications for future HMEs and present qualitative scenarios in Secs. \ref{next_event} and \ref{interpretation}.

\section{Existing monitoring data for Q2237+0305}
\label{data}

\begin{figure}
\vspace{80mm}
\includegraphics{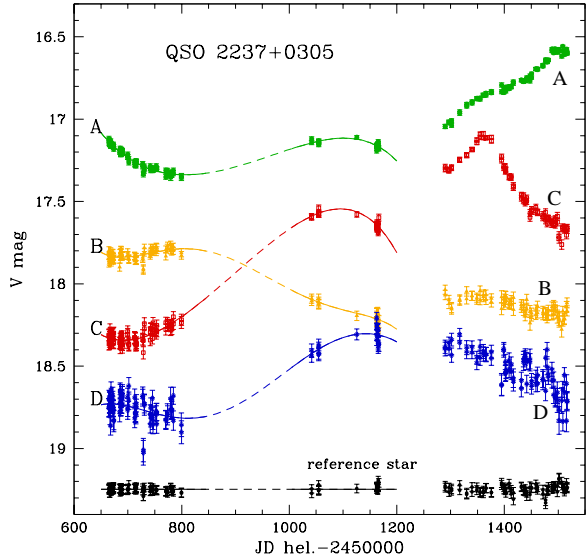}
\caption{\label{OGLE_lc}Image light-curves from the OGLE monitoring data. Figure from the OGLE web page (see http://www.astro.princeton.edu/$\sim$ogle/ogle2/huchra.html).}
\end{figure}

\begin{figure*}
\vspace{100mm}
\includegraphics{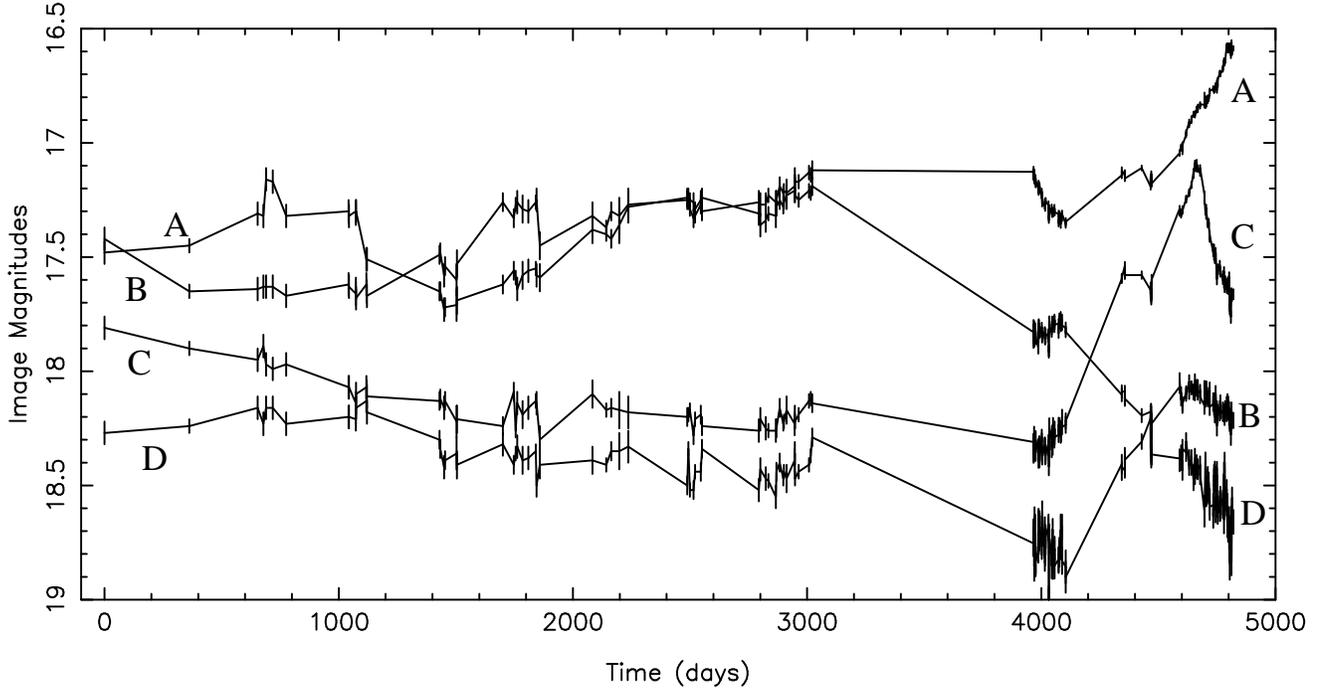}
\caption{\label{comp_lc}Compilation of light-curve data for Q2237+0305.}
\end{figure*}

Fig. \ref{OGLE_lc} shows the OGLE light-curve for Q2237+0305, containing all points taken before the end of the 1999 observing season (figure taken from the OGLE web page (see http://www.astro.princeton.edu/$\sim$ogle/ogle2/huchra.html). Note that this data was published in Wozniak et al. (2000a,b).
 The light-curves have an unprecedented sampling rate ($\sim1$ week). The OGLE data adds to the approximately 10 years of previously obtained, but less densely sampled photometry presented in Schneider et al. (1988), Kent \& Falco (1989), Irwin et al. (1989), Corrigan et al. (1990) and $\O$stensen et al. (1996).
 Fig. \ref{data} shows the complete compiled data set. Error bars are shown representing the published errors. Note that the data taken prior to OGLE is in R and r bands whereas the OGLE monitoring is in V band. The typical amplitudes of microlensing flux variation may therefore differ between data sets if the source size is colour dependent (Wambsganss \& Paczynski 1991).

\section{The Microlensing model}
\label{model}

Our analysis involves the comparison of ensembles of microlensing models to Q2237+0305 light-curve features. Through this comparison we interpret observed HMEs and attempt to predict the features of future HMEs. This section provides a brief discussion of the models used.

 Throughout the paper we use standard notation for gravitational lensing. The Einstein radius of a microlens projected into the source plane is denoted by $\eta_{0} $. The normalised shear is denoted by $\gamma$, and the convergence or optical depth by $\kappa$. The basic model for microlensing at high optical depth comprises a disc of point masses (eg. Kayser, Refsdal \& Stabell 1986) with a size such that a large fraction ($>99\%$) of macroimage flux is recovered (Katz, Balbus \& Paczynski 1986; Lewis \& Irwin 1995). To construct a microlensed light-curve we use the contouring technique of Lewis et al. (1993) and Witt (1993). For the microlensing models of Q2237+0305 used in the current work we assume the macro-parameters for the lensing galaxy calculated by Schmidt, Webster \& Lewis (1998). We will use the standard notation introduced by Yee (1988), to describe these images. Where required a cosmology having $\Omega=1$ with $H_{o}=75\,km\,sec^{-1}$ is assumed. 

We describe the microlensing rate in terms of the effective transverse velocity, which is defined as the transverse velocity that produces a microlensing rate from a static model equal to that of the observed light-curve (Wyithe, Webster \& Turner 1999 (hereafter WWT99)). The effective transverse velocity therefore describes the microlensing rate due to the combination of the effects of a galactic transverse velocity and proper motion of microlenses. To calculate distributions of values for light-curve features associated with HMEs, we assume that the effective transverse velocity accurately describes not only the distribution of light-curve derivatives during an HME (Wyithe, Webster \& Turner 2000a), but also the distribution of orientations between the caustic and source trajectory, and hence the event duration.

We have previously obtained the following normalised probability distributions.
These were obtained under the assumption that the source size $S\ll\eta_o$. Evidence in favour of this assumption was presented in Wyithe, Webster \& Turner (2000c).
 
\noindent $i)$ $p_{s}(S|\langle m \rangle,v_{eff})$, the probability that the continuum source diameter is between $S$ and $S + {\rmn d}S$ given a mean microlens mass $\langle m \rangle$ and an effective galactic transverse velocity $v_{eff}$ (Wyithe, Webster, Turner \& Mortlock 2000).

\noindent $ii)$ $p_{v}(v_{eff}|\langle m \rangle)$ the probability that the effective galactic transverse velocity is between $v_{eff}$ and $v_{eff}+{\rmn d} v_{eff}$ given a mean microlens mass $\langle m \rangle$ (WWT99). 

\noindent $iii)$ $p_{m}(\langle m \rangle)$, the probability that the mean microlens mass is between $\langle m \rangle$ and $\langle m \rangle + {\rmn d} \langle m \rangle$ (Wyithe, Webster \& Turner 2000b).

$p_{v}(v_{eff}|\langle m \rangle)$ and $p_{m}(\langle m \rangle)$ were computed using flat ($p(V_{tran})\propto dV_{tran}$), and logarithmic ($p(V_{tran})\propto \frac{dV_{tran}}{V_{tran}}$) assumptions for the Bayesian prior for galactic transverse velocity ($V_{tran}$). $p_{v}(v_{eff}|\langle m \rangle)$ was found to be insensitive to the prior assumed, however $p_{m}(\langle m \rangle)$ showed some dependence. In the remainder of this paper we use $p_{m}(\langle m \rangle)$ calculated using the assumption of a logarithmic prior. We note that the assumption of the flat prior raises the average light-curve derivative by a few percent.

 The functions $p_{s}(S|\langle m \rangle,v_{eff})$, $p_{v}(v_{eff}|\langle m \rangle)$, $p_{m}(\langle m \rangle)$ and the HME statistics presented in this paper were computed for different assumptions of smooth matter density, photometric error, and direction of the galactic transverse velocity. Since the probability functions referred to above were computed from a derivative analysis, the statistics that we compute in this paper are quantitatively similar for the different possible models. Therefore we present only results from models with no smooth matter, a transverse velocity direction along the image C-D axis and simulated photometric errors assigned according to a Gaussian distribution with half widths of $\sigma=\Delta M/2$ in images A and B, and $\sigma=\Delta M$ in images C and D. 

 Both the microlensing rate due to a transverse velocity (eg. Witt, Kaiser \& Refsdal 1993), as well as the corresponding rate due to proper motions (WWT00a) are not functions of the details of the microlens mass distribution, but rather are only dependent on the mean microlens mass. We therefore limit our attention to models in which all the microlenses have the same mass since the results obtained will be applicable to other models with different forms for the mass function.

The determination of probability for the quantity $v_{eff}\sqrt{\langle m\rangle}$ from the Q2237+0305 monitoring data is quite robust. However the probability for the source size is derived from a single poorly sampled HME. The small number of observations describing the 1988 peak (Irwin et al. 1989; Corrigan et al 1991) introduces the potential for a systematic error in the source size equal to the ratio of the true event length and the inferred event length of $\sim 52$ days (twice the estimated rise time). This can be compared to the $\sim$100 day separation of the two points that provide an upper-bound on the event duration. The resulting systematic error in the estimate of source size is therefore smaller than a factor of $\sim 2$. In addition there may also be a component of systematic error from the assumption that the 1988 peak was due to a single caustic crossing. The statistics presented in the following sections are therefore computed assuming prior probabilities for $S$ assuming no systematic error,
\begin{equation}
p_{s}\left(S|\langle m\rangle,v_{eff}\right)dS,
\end{equation}
and systematic errors of $\times 2$ and $\times 5$ in $S$:
\begin{equation}
p'_{s}\left(S'|\langle m\rangle,v_{eff}\right)dS'=\frac{1}{2}p_{s}\left(2S|\langle m\rangle,v_{eff}\right)dS
\end{equation}
and 
\begin{equation}
p'_{s}\left(S'|\langle m\rangle,v_{eff}\right)dS'=\frac{1}{5}p_{s}\left(5S|\langle m\rangle,v_{eff}\right)dS.
\end{equation}
Our source size estimate was made from data collected in the R and r bands, while the OGLE light-curves showing the features that we wish to investigate are in the V band. This introduces the possibility for another source of systematic error if the source has significantly different sizes in the R/r and V bands.

To investigate individual HMEs we must look at light-curve statistics for single images. Therefore unlike the calculation of $p_{s}$, $p_{v}$ and $p_{m}$, which used difference light-curves, intrinsic source variation may be important. This cannot be directly measured, however in Wyithe, Webster, Turner \& Agol (2000) (hereafter WWTA00) limits are placed on the intrinsic variability power-spectrum and it is shown that intrinsic variability should not be an important consideration during HMEs.

\section{Analysis of the OGLE Light-Curve}

\label{OGLE_lc_sec}

\begin{table*}
\begin{center}
\caption{\label{trigger_probs}Table of probabilities for model light-curve features corresponding to those in the 1999 image C event. Details are discussed in the text.}
\begin{tabular}{|c|c|c|c|c|c|c|}
\hline
Event type & Size Error    &Total fraction & Prob $<2$ weeks & Prob $\dot{M}_{max}<2$  & Prob $M_{height}<0.5$ & Prob $\Delta M_{min}<0$ \\\hline\hline
           & $S\times 1$ & 0.27          & 0.11            &  0.83                   &   0.55                & 0.50                     \\
Cusp       & $S\times 2$ & 0.27          & 0.19            &  0.84                   &   0.55                & 0.50                     \\
           & $S\times 5$ & 0.26          & 0.12            &  0.87                   &   0.57                & 0.50                     \\\hline

           & $S\times 1$ & 0.14          & 0.07            &  0.01                   &   0.02                & 0.08                     \\
$+ve$ caustic & $S\times 2$& 0.18          & 0.06            & 0.01                    &  0.02                 & 0.08                     \\
           & $S\times 5$ & 0.32          & 0.09            &  0.10                   &   0.03                &  0.10                    \\\hline
 
           & $S\times 1$ & 0.58          & 0.05            & 0.04                    &  0.13                 & 0.98                     \\
$-ve$ caustic& $S\times 2$& 0.55          & 0.04            & 0.12                    & 0.20                  & 0.98                     \\
           & $S\times 5$ & 0.41          &  0.02           &  0.45                   &  0.30                 &  0.98                    \\\hline

\end{tabular}
\end{center}
\end{table*}

 In this section we apply simple statistics describing event heights and light-curve derivatives a postiori to specific features in the published light-curves. We also retrospectively apply the triggering function developed in WWTA00 to two observed light-curve peaks. We use the results to interpret features found in monitoring data. In particular we discuss whether the large scale variation in image C (1999) was due to the source having crossed a caustic or moved outside of a cusp. 

We note that choosing light-curve features a-postiori, and comparing them to a sample of models in order to draw conclusions regarding the type of event involved has the potential to introduce a statistical bias. However in the present case we feel that our method is justified because of the well established prior knowledge that caustic crossing HMEs produce larger amplitude, shorter period and more asymmetric (in time) variation than cusp related events.

\subsection{Statistics of the 1999 light-curve Peak in image C}

The image C light-curve shows a remarkable resolved peak described by $\sim 35$ points on separate days spanning $\sim 7$ months (see Fig. \ref{OGLE_lc}).
By December 1999 the image C light-curve had dropped to a level similar to the end of the 1998 observing season. The peak is quite symmetric having reached a height $\sim 0.5$ magnitudes above the 1998 level. The maximum derivative reached both before and after the event peak was $\sim 2$ mags/year. Images B and D remained fairly constant during this period suggesting constant intrinsic luminosity. The event properties are discussed in relation to model calculations of the probability for their values. The results are summarised in Tab. \ref{trigger_probs}.

\subsubsection{Maximum light-curve derivative}

\begin{figure}
\vspace{65mm}
\includegraphics{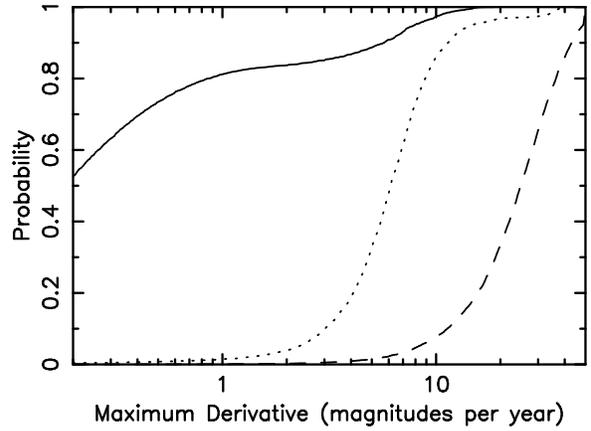}
\caption{\label{max_deriv}Plots of the probability for the maximum derivative preceding each a light-curve peak. Functions are given corresponding to derivatives that precede cusps (solid line), and caustic crossings with disappearing (dotted line) and appearing (dashed line) critical images. Only peaks larger than 0.1 magnitudes above the local minima on both sides are considered.  }
\end{figure}

We have calculated the cumulative probability of observing a maximum light-curve derivative $\dot{M}$ prior to the peak maximum: 
\begin{eqnarray}
\nonumber
&&\hspace{-7mm}P_{\dot{M}}(\dot{M}<\dot{M}_o)=\\
\nonumber&&\hspace{-3mm}\int dm\int dS\int dv_{eff}\,\,\left( p_{s}\left(S|\langle m\rangle,v_{eff}\right)\,p_{m}\left(\langle m\rangle\right)\right.\\
&&\hspace{5mm}\times\left.p_{v}\left(v_{eff}|\langle m\rangle\right)P_{\dot{M}}\left(\dot{M}<\dot{M}_o|S,\langle m\rangle,v_{eff}\right)\right),
\end{eqnarray}
using a sampling rate typical of the monitoring data (1 point per week). We have looked at the probability $P$ for cusps ($P_{\dot{M}_C}(\dot{M}<\dot{M}_o)$), $-ve$ caustic crossings ($P_{\dot{M}_-}(\dot{M}<\dot{M}_o)$) and $+ve$ caustic crossings ($P_{\dot{M}_+}(\dot{M}<\dot{M}_o)$). 
Fig. \ref{max_deriv} shows these functions; solid, dotted and dashed lines correspond to $P_{\dot{M}_C}(\dot{M}<\dot{M}_o)$, $P_{\dot{M}_-}(\dot{M}<\dot{M}_o)$, and $P_{\dot{M}_+}(\dot{M}<\dot{M}_o)$ respectively. The functions shown assume that the source size estimate $S$ is correct.
We find that a maximum derivative of $\sim 2$ magnitudes per year is inconsistent at $>99\%$ level with the event having been a $+ve$ caustic crossing. In addition, we find that a $-ve$ caustic crossing is excluded at the 95\% level. A cusp cannot be excluded on the basis of the maximum derivative (though it is higher than expected). Our conclusions are barely changed (the $-ve$ event confidence is then 90\%) if we assume our source size has been underestimated by a factor of 2. If our underestimate is a factor of 5 then only the $+ve$ event can be excluded (at the 90\% level). The results are summarised in column 5 of Tab. \ref{trigger_probs}.

\subsubsection{Height of the peak above the previous local minimum}

\begin{figure}
\vspace{65mm}
\includegraphics{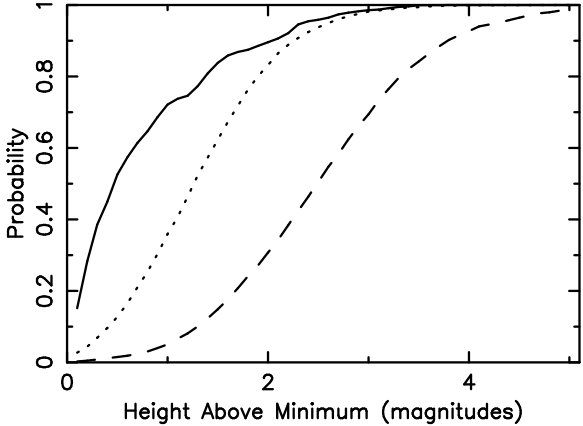}
\caption{\label{max_height}Plots of the probability for the difference between a light-curve peak maximum and the preceding minimum. Functions are given corresponding to the peak heights for cusps (solid line), and caustic crossings with disappearing (dotted line) and appearing (dashed line) critical images. Only peaks larger than 0.1 magnitudes above the local minima on both sides are considered. }
\end{figure}

If the light-curve minimum preceding the 1999 image C event is assumed to have been at the level of the 1998 season then the height of the peak maximum above the previous local minimum is $M_{peak}\sim 0.5$ magnitudes. We calculate the cumulative probability for the difference between peak maxima and the preceding local minima ($M_{height}$). $M_{height}$ is a function of sampling rate for sharp peaks and sparse samplings. We have therefore used a sampling rate of 1 point per week and integrated the probability over $v_{eff}$ as well as $S$ and $\langle m\rangle$:
\begin{eqnarray}
\nonumber
&&\hspace{-7mm}P_{M}(M_{height}<M_{peak})=\\
\nonumber&&\hspace{-3mm}\int dm\int dS\int dv_{eff}\,\,\left( p_{s}\left(S|\langle m\rangle,v_{eff}\right)\,p_{m}\left(\langle m\rangle\right)\right.\\
&&\hspace{0mm}\times\left.p_{v}\left(v_{eff}|\langle m\rangle\right)P_{M}\left(M_{height}<M_{peak}|S,\langle m\rangle,v_{eff}\right)\right).
\end{eqnarray}

\noindent As before we have computed the function for cusps ($P_{M_{C}}(M_{height}<M_{peak})$), $+ve$ caustic crossings ($P_{M_{+}}(M_{height}<M_{peak})$), and $-ve$ caustic crossings ($P_{M_{-}}(M_{height}<M_{peak})$). Fig. \ref{max_height} displays the resulting curves (assuming no systematic uncertainty in the source size estimate). $\Delta M_{peak}\sim 0.5$ magnitudes is typical if the event is due to a cusp, and is ruled out at the 98\% level if the event is $+ve$ caustic crossing. If the event is a $-ve$ caustic crossing the results are inconclusive. A very similar conclusion is reached if the source size has been underestimated by a factor of 2. The results are summarised in column 6 of Tab. \ref{trigger_probs}.

\subsubsection{A postiori application of the triggering function to event 1999C}

\label{trigger_Im_C}

\begin{figure}
\vspace{65mm}
\includegraphics{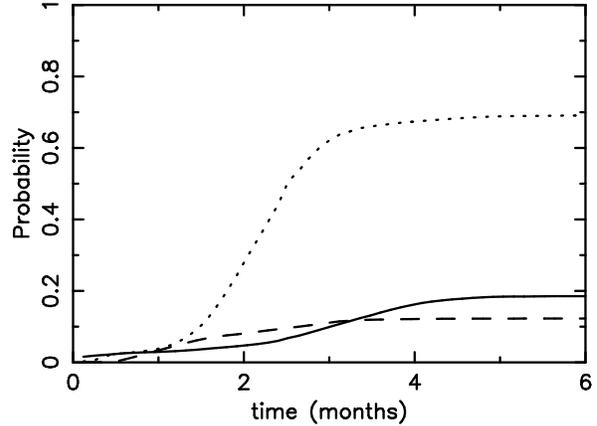}
\caption{\label{obs_trigger}Plots of the probability for the delay of a light-curve peak following the observed image C light-curve derivative on the 19th June. Functions are given corresponding to the delays for cusps (solid line), and caustic crossings with disappearing (dotted line) and appearing (dashed line) critical images.}
\end{figure}

WWTA00 describes a general function to determine how long one should wait ($P$) for a HME following a hypothetical observed light-curve derivative ($T$). If a light-curve derivative $T_{obs}\pm\Delta T_{obs}$ is observed, predictions  $F(P|T_{obs}\pm\Delta T_{obs})$ can be made about forthcoming events specifically for the current data using the sampling rate identical to that of the observations. On the 19th of June 1999 monitoring from the OGLE collaboration (OGLE web page) showed image C rising at a rate of 1.21-1.78 mags/year. Since at this level $P$ is not sensitive to the sampling rate for sampling spacings smaller than two weeks, we used observations on the 10th June, 19th June and 1st July.

For the present calculation the triggering function discussed in WWTA00 is slightly modified. Rather than searching for peaks following a derivative larger than some hypothetical value, we search for peaks following light-curve derivatives in the measured range. Thus since the algorithm steps along the curve in steps equal to the sampling rate, certain events (in particular $+ve$ caustic crossings) are missed if the derivative jumps from below the lower bound to above the upper bound during one sampling interval. Therefore the results describe the relative likely-hood of observing the different types of event following a current light-curve derivative. As a consequence of the triggering algorithm searching for derivatives along the light-curve in one direction, then the model light-curve derivative at the point of the trigger is systematically biased toward $T-\Delta T$. $P$ is therefore an overestimate. This bias is minimised by reduced photometric error.

 Fig. \ref{obs_trigger} shows the resulting triggering functions $F_{+}$, $F_{-}$ and $F_{C}$ computed assuming no systematic uncertainty in the source size estimate. The solid, dotted and dashed lines correspond to $F_{C}(P|T_{obs}\pm\Delta T_{obs})$, $F_{-}(P|T_{obs}\pm\Delta T_{obs})$, $F_{+}(P|T_{obs}\pm\Delta T_{obs})$ respectively. We find that the observed trigger precedes a caustic crossing HME $\sim 75\%$ of the time. The event peak is most likely to occur $\sim 3$ months following such a trigger if the event was a cusp, 2 months later if it was a $-ve$ caustic and 1 month later if it was a $+ve$ caustic crossing. Note that having observed a derivative in the quoted range means that a $+ve$ caustic crossing is very unlikely due to the rapid change in the light-curve derivative. The light-curve peaked at $\sim 1$ July, about 2-weeks after the observed derivative. This is surprisingly early for all types of events (though $P$ is an overestimate). At the $90\%-95\%$ level, caustic crossing events are excluded. This result holds if a systematic source size error is assumed. The probabilities of both the class of event following the trigger as well as the arrival time are summarised in columns 3 and 4 of Tab. \ref{trigger_probs}.

\subsection{What class of event have we seen in image C ?}

Constraints on the event type of the OGLE image C HME are placed by both the maximum derivative observed prior to the event peak, and the height of the peak above the previous minima. $+ve$ caustic crossings are excluded by both these measurements (even when potential systematic errors in source size are assumed). In addition, assuming no systematic uncertainty in source-size, $-ve$ caustic crossings are excluded by the measured maximum derivative. The cusp interpretation is consistent with both measurements.

Because triggers in the 1999 image C light-curves trigger were relatively small, the fraction of events predicted for different classes of HME by the triggering function does not restrict the type of event observed. In particular, the trigger was not large enough to rule out either cusps or $-ve$ caustic crossings. However the OGLE data shows a previous rise in the light-curve of image C occurring between $\sim 300$ and $500$ days prior to July 1999. This rise occurred between observing seasons, but the net change is $\ga0.8$ magnitudes suggesting that a $+ve$ caustic crossing event may have been missed (see Sec. \ref{97-98_C}). The typical separation of double peaked events calculated by Witt, Kayser and Refsdal (1993) is 300-500 days. This fact coupled with the comparative rarity of a $+ve$ event following the observed trigger leads to the conclusion that if the 1999 event were a caustic crossing then it is more likely to have been the second peak in a double horned event ($-ve$ caustic crossing) than a $+ve$ caustic crossing. However, based on the peak arrival time, a cusp event is more likely than a $-ve$ caustic crossing (or a $+ve$ caustic crossing).

 Based on this evidence we conclude that the event observed for image C was probably a cusp event rather than a caustic crossing. However, further monitoring may show that this interpretation is false, and therefore that the microlensing models are in error. Thus the event will provide a valuable test of whether our current microlensing models can be used to reliably interpret the initial stages of HME's.

\subsubsection{Asymmetry of the 1999 image C event}

\begin{figure}
\vspace{65mm}
\includegraphics{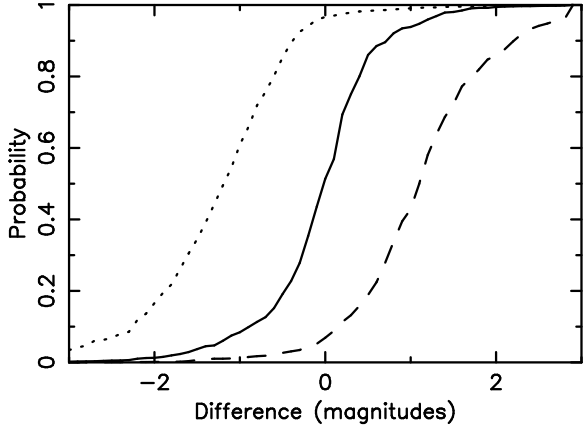}
\caption{\label{assym}Plots of the probability for the difference between the local light-curve minima either side of the peak maximum. Functions are given corresponding to the probabilities for cusps (solid line), and caustic crossings with disappearing (dotted line) and appearing (dashed line) critical images. Only peaks larger than 0.1 magnitudes above the local minima on both sides are considered. }
\end{figure}

 At the time of writing the image C light-curve was still in decline, but appeared to be decelerating ($+ve$ second derivative). Observations of the final stages of this event will provide a further property that can be used to distinguish between the different possibilities. After the light-curve has flattened out, the difference between the light-curve minima immediately preceding and following the event can be measured: 
\begin{equation}
\Delta M_{min} = M_{min}(left)-M_{min}(right). 
\end{equation}
We have calculated the distribution of these values as before:
\begin{eqnarray}
\nonumber
&&\hspace{-7mm}P_{\Delta M_{min}}(\Delta M_{min}<\Delta M_{obs})=\\
\nonumber&&\hspace{-5mm}\int dm\int dS\int dv_{eff}\,\,\left( p_{s}\left(S|\langle m\rangle,v_{eff}\right)\,\right.\\
\nonumber&&\hspace{0mm}\times p_{m}\left(\langle m\rangle\right)\,p_{v}\left(v_{eff}|\langle m\rangle\right)\\
&&\hspace{3mm}\times\left.P_{\Delta M_{min}}\left(\Delta M_{min}<\Delta M_{obs}|S,\langle m\rangle,v_{eff}\right)\right).
\end{eqnarray}
The functions were calculated corresponding to cusps, $-ve$ caustic crossings and $+ve$ caustic crossings ($P_C$, $P_-$ and $P_+$ respectively). The resulting curves are plotted in Fig. \ref{assym} (for the case of no systematic error in the source size estimate). The results are summarised in column 7 of Tab. \ref{trigger_probs}.

While the trailing minimum has not yet been observed, it may provide a discriminate between the interpretations of the 1999 image C event as a cusp and a $-ve$ caustic. In particular, if the light-curve flattens out at a level approximately equal to that of the 1998 season ($\Delta M_{min}\sim 0$), then the $-ve$ caustic crossing interpretation will be ruled out at the 95\% level. Assuming that the previous light-curve minimum occurred during 1998, the $+ve$ caustic crossing interpretation is already ruled out at $>95\%$.

\subsection{Comparison with the 1998 image A event}

\begin{figure}
\vspace{65mm}
\includegraphics{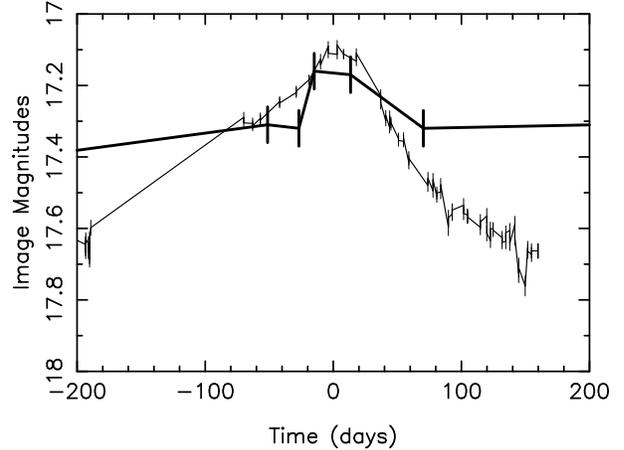}
\caption{\label{peak_compare} The light-curves for the 1988 image A peak (thick line), and the 1999 image C peak (thin-line). The two peaks have been placed on the same time-axis such that the peaks approximately coincide at $t=0$.}
\end{figure}

\begin{figure}
\vspace{65mm}
\includegraphics{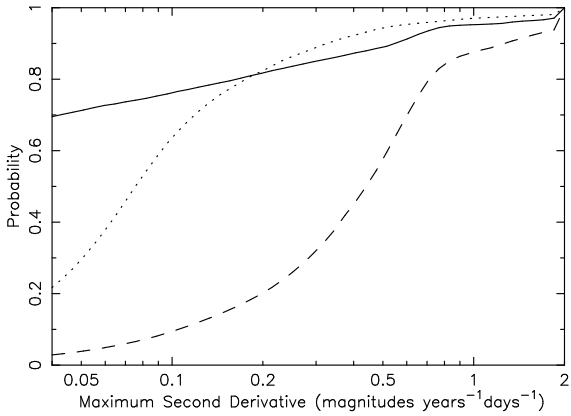}
\caption{\label{second_deriv}Plots of the probability for the maximum second derivative on the leading side of the peak maximum. Functions are given corresponding to the probabilities for cusps (solid line), and caustic crossings with disappearing (dotted line) and appearing (dashed line) critical images. Only peaks larger than 0.1 magnitudes above the local minima on both sides are considered. }
\end{figure}

\begin{figure*}
\vspace{70mm}
\includegraphics{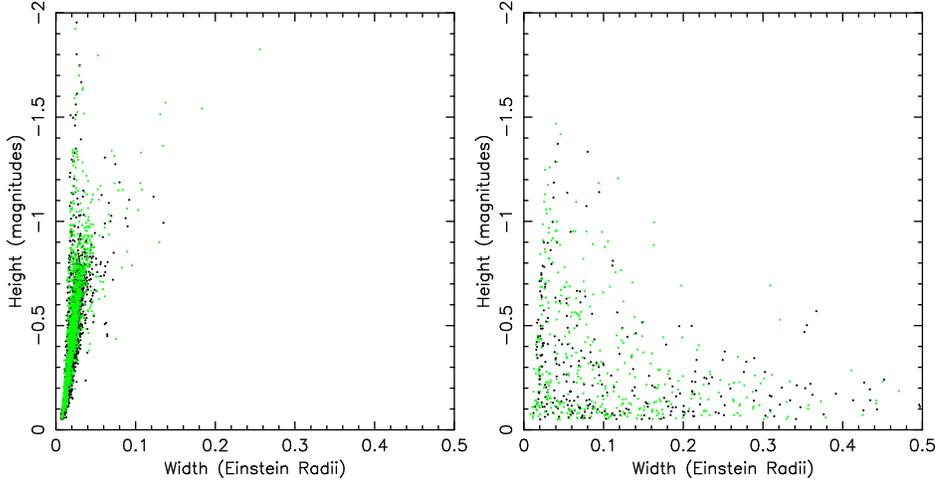}
\caption{\label{height_v_width} Scatter plots of peak-height above half maximum vs. full width at half maximum. Left: peaks due to caustic crossings. Right: peaks due to cusp events. Light and dark dots correspond to images A and C respectively. $\gamma_A>0$, $\gamma_C<0.$} 
\end{figure*}

The first unambiguous microlensing signal was the rise and fall of image A in 1988 during a period when the other images remained at a consistent level (ie. no intrinsic variability). Some insight is gained through comparison of the shape of this light-curve peak with the 1999 image C event. The two events are described by very different data sets. While the OGLE data has provided excellent coverage of the 1999 peak, allowing quantities such as the maximum derivative and peak height to be calculated, the 5 observations of the 1988 peak provide a much cruder record. In particular the values of maximum pre-peak derivative, the peak height above the minimum and the difference between the minimum cannot be computed. 
 On the other hand, our source size estimation was determined from the assumption that the image A peak is a caustic crossing. Statistical determinations of the event type are therefore only meaningful as a check of the self-consistency of the calculations. In that vein we note that the one leading derivative that can be measured from the data is $T\sim5$ mags/year. Calculations in WWTA00 suggest that the observations should have been followed by a caustic crossing in $\sim2-3$ weeks or a cusp event in $\sim5$ weeks, which lie between and after the two brightest observations respectively.  

Fig. \ref{peak_compare} shows light-curves of the two peaks, placed on the same time-axis such that the maxima approximately coincide at $t=0$.
The image A and C peaks are shown by thick and thin lines respectively. The large initial rise of the 1988 event measures the minimum to the maximum gradient (which is surely significantly larger given the high second derivative). This sharp rise is not replicated in the 1999 event which had a maximum derivative of $\sim2$ magnitudes per year. Since we have a lower bound for the maximum derivative of the 1988 image A event which is consistent with all three types of HME, the size of the observed derivative cannot be used as a discriminate (although for a cusp event there is a 90\% chance of $\dot{M}_{max}<5$ magnitudes per year). However, the leading three points also measure a lower limit for the maximum second derivative. Fig. \ref{second_deriv} shows probability $P_{\ddot{M}}$ for the maximum value of second derivative on the leading sides of cusp events (solid line) and $+ve$ (dashed line) and $-ve$ (dotted line) caustic crossings computed for image A (no systematic error in source size was assumed):
\begin{eqnarray}
\nonumber
&&\hspace{-7mm}P_{\ddot{M}}(\ddot{M}<\ddot{M}_{obs})=\\
\nonumber&&\hspace{-3mm}\int dm\int dS\int dv_{eff}\,\,\left( p_{s}\left(S|\langle m\rangle,v_{eff}\right)\,p_{m}\left(\langle m\rangle\right)\right.\\
&&\hspace{5mm}\times\left.p_{v}\left(v_{eff}|\langle m\rangle\right)P_{\ddot{M}}\left(\ddot{M}<\ddot{M}_{obs}|S,\langle m\rangle,v_{eff}\right)\right).
\end{eqnarray}
The derivatives were calculated using a sampling rate corresponding to the initial 3 observations of the 1988 peak. Comparing the image A second-derivative of $\ddot{M}_{obs}\sim 0.5$ magnitudes per year per day to Fig. \ref{second_deriv} we infer that the 1988 peak was probably a $+ve$ caustic crossing.

A second feature to be noted from Fig. \ref{peak_compare} is that the 1999 peak appears to have a much longer duration than the 1988 peak. Fig. \ref{height_v_width} shows scatter plots of event height above full-width-at-half-maximum (fwhm) vs. fwhm for images A (light dots) and C (dark dots). The plot on the left shows the relationship for caustic crossings while the right-hand plot is for cusp events. The plots highlight the rise-time - peak-height correlation for caustic crossings, and the cloud of points due to smooth light-curve variations (both of which were pointed out by Witt \& Mao (1994)). The separation of the points into two categories demonstrates the intuitive notion that cusp events have longer durations than caustic crossings. The cusp event parameters shows a peak width lower-limit corresponding to the caustic crossing correlation. Therefore, the systematic bias introduced into the source size determination by the assumption that the 1988 event was due to a caustic crossing can only result in an over estimate of source size. In addition, Fig. \ref{height_v_width} demonstrates that a cusp event contains very little information on source size/transverse velocity due to the lack of any correlation with peak height. The inference that the 1988 peak was a caustic crossing and the 1999 peak was a cusp event is consistent with Fig. \ref{height_v_width}.

\subsection{A 1997-1998 event in the OGLE image C light-curve ?}
\label{97-98_C}
Fig. \ref{data} shows a $\Delta M_{obs}\sim 0.8$ magnitude rise between the 1997 image C minimum and the 1998 level, suggesting an event in between those observing seasons. We assume that the intrinsic source luminosity was approximately constant over this period, which is supported both by the facts that image A changed by $\la0.2$ magnitudes and that the other images show opposite trends. Comparing the image C change to the probabilities in Fig. \ref{assym} we find that $\Delta M_{obs}$ is consistent with a $+ve$ caustic crossing having occurred between the 1997 and 1998 observing seasons, but rules out $-ve$ caustic crossings ($\sim 99\%$) and cusp events ($\sim 95\%$). We therefore infer that a $+ve$ caustic crossing was missed between the 1997 and 1998 observing seasons.

\section{Predictions of future HMEs}
\label{next_event}

In this section we calculate probability functions for the likelihood of observing future HMEs in images A and C given current light-curves.

\subsection{The next image C HME}
\label{next_event_C}

\begin{figure}
\vspace{137mm}
\includegraphics{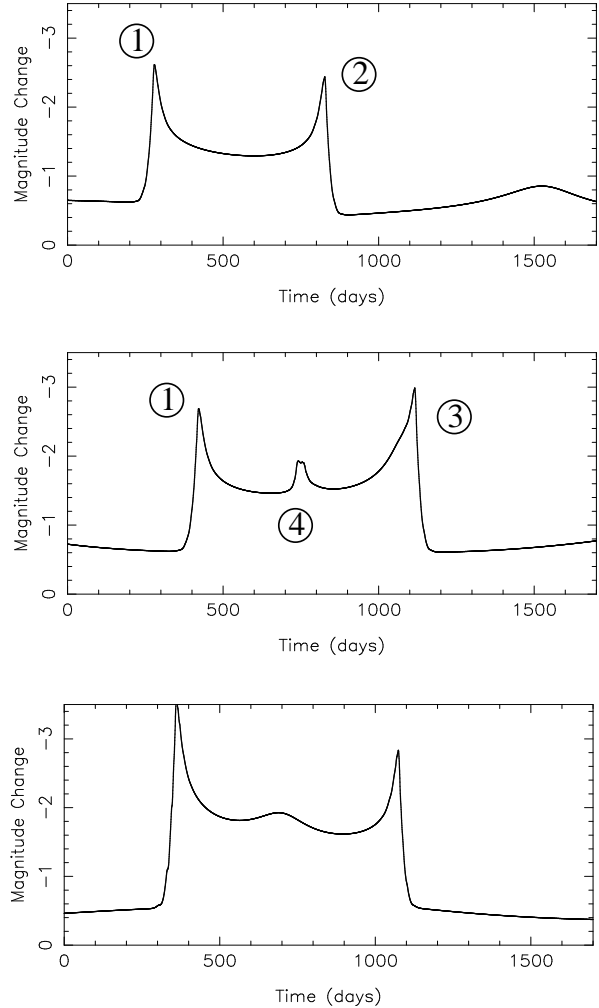}
\caption{\label{scenarios_real}Examples of double-horned profile (top) and double horned profile with an additional cusp event (centre and bottom).}
\end{figure}

\begin{figure*}
\vspace{70mm}
\includegraphics{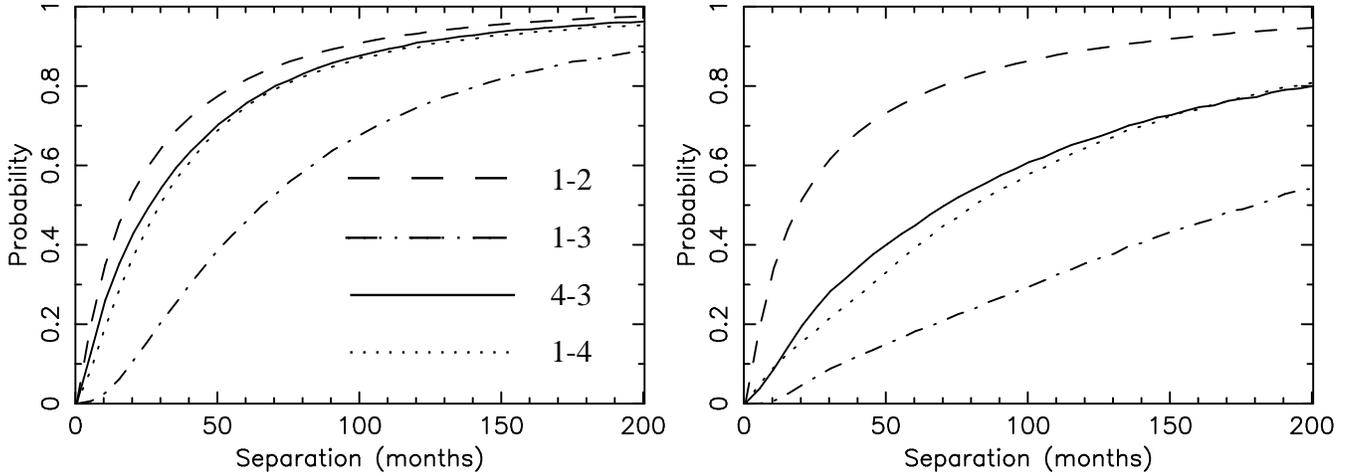}
\caption{\label{caust_prob}Solid lines: Plots of the probability for the time in months until the next caustic crossing assuming that the 1999 light-curve peak is a cusp related event following a $+ve$ caustic crossing. Dashed lines: Plots of the probability for the separation in months of a $+ve$ HME followed by a second HME. Dot-Dashed lines: Plots of the probability for the separation in months of a $+ve$ HME and a second HME that bracket a cusp event. Dotted lines: probability for the separation of a cusp event following a $-ve$ caustic crossing. The left and right hand plots correspond to transverse velocity directions aligned with the C-D and A-B axes respectively. Only peaks larger than 0.1 magnitudes above the local minima on both sides are considered.}
\end{figure*}

\begin{figure}
\vspace{65mm}
\includegraphics{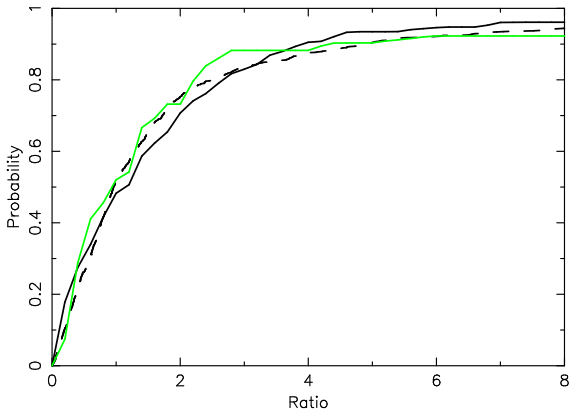}
\caption{\label{caust_ratio} The probability for the ratio of separations in months between a cusp peak (following a $-ve$ caustic crossing and the nearest caustic on either side. The dark and light-lines correspond to transverse velocity directions aligned with the C-D and A-B axes respectively. Only peaks larger than 0.1 magnitudes above the local minima on both sides are considered.} 
\end{figure}

In this section we assume that there was a $+ve$ caustic crossing between the 1997 and 1998 observing seasons, and investigate when we should next see a caustic crossing in image C. These calculations follow Witt, Kayser \& Refsdal (1993) who calculated the separations in dimensionless units of the different combinations of $+ve$ and $-ve$ events. However we have included both our estimates of $\langle m\rangle$ and $v_{eff}$, and the cusp as a third class of HME.

Due to the typical diamond formation of fold caustics, the case of a $+ve$ followed by a $-ve$ caustic crossing is common. Similarly, inspection of model light-curves shows that cusp events follow $-ve$ caustic crossings as the source moves past the cusp associated with that caustic (this feature is seen in the double horned profile that is characteristic of the Chang-Refsdal lens). However we have inferred that the OGLE image C light-curve shows a $+ve$ caustic crossing followed by a cusp event. Such a combination is much less common and is due to the source moving past a cusp formed from two fold caustics other than the one responsible for the $+ve$ caustic crossing HME. It can also be seen in model light-curves. Examples of the two scenarios are shown in Fig. \ref{scenarios_real}. The upper panel shows an example of a double horned event. The lower two panels show examples of a double horned event surrounding a cusp event. The source is shown passing very close to the cusp (centre panel), partly coming in contact, and passing further away, producing a lower amplitude event (lower panel). The light-curves were produced using our most likely model for the microlensing parameters ($v_{eff}=300km\,sec^{-1}$, $\langle m\rangle=0.1\,M_{\odot}$ and $S=5\times10^{14}\,cm$). The intervals (eg. 1-2) quoted below refer to the intervals between the events labelled on these plots.

From these two scenarios, we generate the probability functions for 4 different event separation statistics (plotted in Fig. \ref{caust_prob}). The left hand and right hand panels show functions computed for transverse velocities aligned with the C-D and A-B axes respectively. Fig. \ref{caust_prob} also contains a key corresponding to the intervals between event types shown in Fig. \ref{scenarios_real}. These intervals are defined below. 

\noindent Firstly, we assume that the 1999 peak was a $-ve$ caustic crossing.

\noindent $i)$ $interval$ 1-2: The dashed lines show cumulative probabilities for the separation of two adjacent caustic crossing HMEs where the first is a $+ve$ caustic crossing. 

\noindent Secondly, the 1999 peak is interpreted as a cusp event following a $+ve$ caustic crossing. 

\noindent $ii)$ $interval$ 1-3: The dot-dashed line shows the distribution of caustic separations in a double horned event where a cusp lies between the two caustics. In this case the typical separation of caustics is larger due to the fact that the cusp is generally formed from independent caustics, so that the chance of a cusp lying inside a diamond or another cusp is higher if the separation (and therefore area) is larger. 

\noindent $iii)$ $interval$ 1-4: The dotted lines represent the probability for the separation between a cusp and the (immediately) preceding $+ve$ caustic crossing.

\noindent $iv)$ $interval$ 4-3: The solid lines represent the probability for the separation between a cusp event (that has followed a $+ve$ caustic crossing) and the subsequent caustic crossing.

\noindent As expected $iii)$ and $iv)$ are very similar. The separation of an inferred $+ve$ caustic crossing between the 1997 and 1998 seasons and the 1999 image C peak is consistent with Fig. \ref{caust_prob} whether the 1999 peak is interpreted as a $-ve$ caustic crossing or a cusp event. However if the 1999 peak is interpreted as a cusp, we expect another caustic crossing to follow. Fig. \ref{caust_prob} shows that in this case, the separation between the cusp peak and the subsequent caustic is likely to be $\ga5$ years (although may be significantly longer).

These statistics for the separation of caustic crossings that surround a cusp does not make use of the information that the observed cusp event followed the caustic crossing by $\sim 500$ days. An alternative way of analysing the separation between the cusp event peak and the next caustic crossing is to calculate the ratio of the time between the cusp event peak and last caustic crossing and the time between the cusp event peak and next caustic crossing. The probability of this ratio is plotted in Fig. \ref{caust_ratio}. The typical ratio is 1 and we expect it to be between $\sim \frac{1}{4}$ and $\sim4$, yielding a most likely arrival time for the next event of $\sim 500$ days, and an upper limit of $\sim 2000$ days ($\sim90\%$). Also plotted in Fig. \ref{caust_ratio} is the probability of the ratio for a random position (dashed lines). The agreement of the curves illustrates the independence of the positions of the cusps and unassociated caustics.

\subsection{The next image A HME}
\label{next_event_A}

\begin{figure}
\vspace{65mm}
\includegraphics{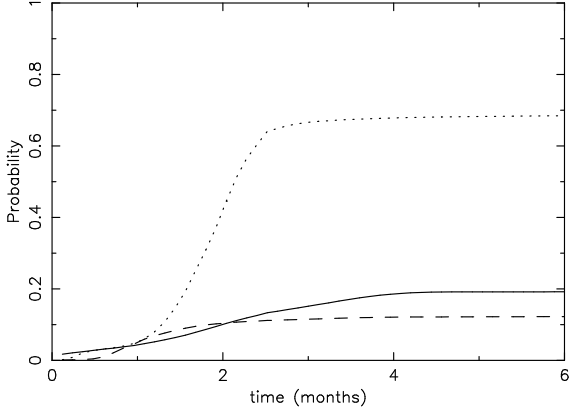}
\caption{\label{obs_trigger_A}Plots of the probability for the delay of a light-curve peak following the observed image A light-curve derivative on the 30th October. Functions are given corresponding to the delays for cusps (solid line), and caustic crossings with disappearing (dotted line) and appearing (dashed line) critical images.}
\end{figure}

\begin{figure}
\vspace{65mm}
\includegraphics{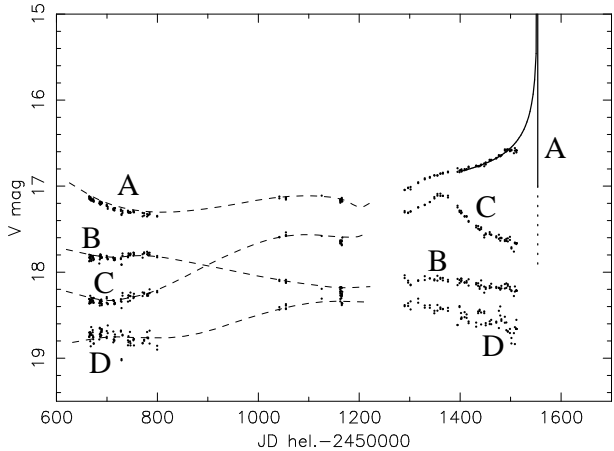}
\caption{\label{fit}Best fit of the near caustic approximation (Eqn. \ref{near_caust}) to the rise in the image A light-curve.}
\end{figure}

In addition to the remarkable peak observed during 1999 in the image C light-curve, the OGLE data also shows image A brightening over the entire season, with the most rapid variations occurring in the latter observations. We have applied the triggering function (as described in Sec. \ref{trigger_Im_C}) to the image A light-curve. 

On the 30th of October 1999 monitoring by OGLE (OGLE web page) showed rises in image A, of 1.41-1.88 mags/year (we used observations on the 20th of October, 30th of October and 9th of November). Fig. \ref{obs_trigger_A} shows the resulting triggering functions $F_{+}$, $F_{-}$ and $F_{C}$. Functions are shown assuming that the source size estimate $S$ is correct. The solid, dotted and dashed lines correspond to $F_{C}(P|T_{obs}\pm\Delta T_{obs})$, $F_{-}(P|T_{obs}\pm\Delta T_{obs})$, $F_{+}(P|T_{obs}\pm\Delta T_{obs})$ respectively. We find that the results are similar to those obtained for the June image C trigger. The observed trigger precedes a caustic crossing HME $\sim 80\%$ of the time. The event peak is most likely to occur $\sim 1-3$ months following such a trigger if the event was a cusp, $\sim 2$ months later if it was a $-ve$ caustic and $\sim 1$ month later if it was a $+ve$ caustic crossing. Having observed a derivative in the quoted range means that a $+ve$ caustic crossing is very unlikely, and a $-ve$ caustic crossing is the most likely option. Unfortunately, these results predict that an event will occur in image A between the 1999 and 2000 observing seasons. If the impending event is assumed to be a $-ve$ caustic crossing, with a previous minimum occurring during the 1998 season, then from Fig. \ref{assym} we predict that the image A light-curve should have a subsequent minimum at a level $\sim 1-1.5$ magnitudes fainter than the November 1999 level. 

If the source is small with respect to $\eta_o$ and therefore the inter-caustic spacing, and the brightening of image A is due to the imminent disappearance of a pair of critical images, then the rise can be modelled using the near caustic approximation of Chang and Refsdal (1979): The flux $f_p$ of a point source at a small time $\Delta t = t_{caust}-t$ from a fold caustic is 
\begin{equation}
\label{near_caust}
f_p = f_o + \theta(\Delta t)\frac{a_o}{\sqrt{\Delta t}},
\end{equation}
where $f_o$ is the magnification of the non-critical images, $\theta(\Delta t)$ the Heaviside step function and $a_o$ a constant describing the strength of the caustic.
The choice of which points to use in a fit of this type is somewhat arbitrary. We have chosen the data following JD 1400, which is after the apparent inflection in the light-curve.
The fit is shown in Fig. \ref{fit}, giving parameter values of $f_o=0.40\,mJy$, $a_o=3.50\,mJy\,days^{\frac{1}{2}}$ and $t_{caust}=1554\,days$ ($\sim$ 11th January 2000). The final figure is of particular value since it predicts the time of the caustic crossing. $t_{caust}$ agrees with the most popular value for the caustic arrival time according the triggering calculation.	

\section{Light-curve interpretations}

\label{interpretation}

\begin{figure}
\vspace{90mm}
\includegraphics{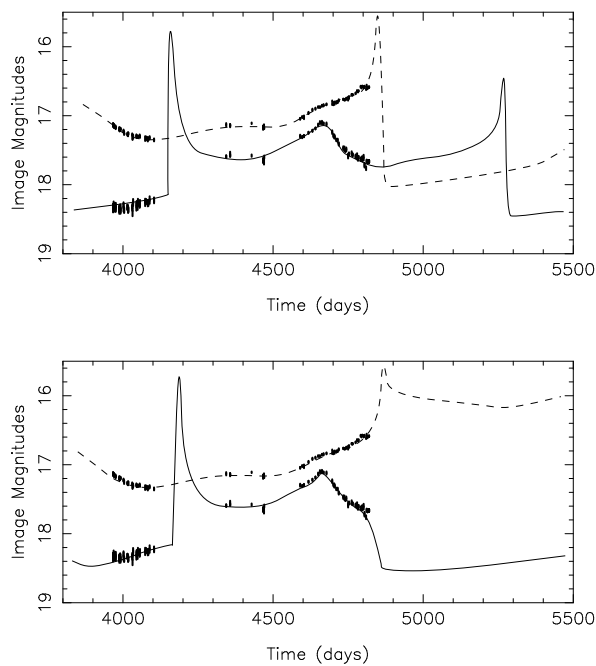}
\caption{\label{scenarios}Cartoons of our interpretations of past light-curves and predictions of future behaviour.}
\end{figure}

While it is pointless to try and fit the observed light-curves directly either by a functional form (except very near a caustic crossing), or with model light-curves, it is illustrative in summary to draw by hand possible interpretations of the data. Fig. \ref{scenarios} shows data for images A and C from Wozniak et al. (2000a,b) and the OGLE web page. The top panel shows possible light-curves corresponding to our most probable interpretation. The solid curve shows the inferred image C event at $\sim 4200$ days, the 1999 light-curve peak as a cusp, and the potential second caustic crossing discussed in Sec. \ref{next_event_C}. The dashed line shows the potential $-ve$ caustic crossing expected following the rise in the light-curve at the end of the 1999 observing season. The lower panel in Fig. \ref{scenarios} shows schematics of less favoured options: the interpretation of the 1999 image C peak as a $-ve$ caustic crossing, and the rise in image A heralding a $+ve$ caustic crossing. 

\section{conclusion}

We have applied simple event statistics a postiori to features in the OGLE light-curves of Q2237+0305. In the specific case of the 1999 peak we conclude that the event was due to the source passing outside of, but close to a cusp rather than to a caustic crossing. This hypothesis may be confirmed or refuted when the trailing peak minimum is observed during the 2000 observing season. In addition, we find that the image C light-curve rise between the 1997 and 1998 OGLE observing seasons was a caustic crossing that resulted in two new critical images. The hypothesis of a cusp event following the first half of a double horned event is a rare feature. However its consequence is that we expect another caustic crossing high magnification event to follow in the image C light-curve. Our models predict that this caustic crossing is most likely to arrive in $\sim 500$ days, and can be expected within $\sim 2000$ days ($\sim 90\%$ confidence). However it may be considerably longer before the next image C caustic crossing is observed, particularly if the transverse velocity moves perpendicular to the shear vector in image C. By applying the triggering function developed in WWTA00 to the rise in the image A OGLE light curve, we predict that a caustic crossing high magnification event will occur between the 1999 and 2000 observing seasons which will result in the loss of a pair of critical images. We therefore expect that the image image A light-curve should have a minimum at a level $\sim 1-1.5$ magnitudes fainter than November 1999.

\section{acknowledgements}
The authors would like to thank Przemyslaw Wozniak for many useful discussions. We would also like to acknowledge the OGLE collaboration for making their monitoring data publically available before publication. This work was supported by NSF grant AST98-02802. JSBW acknowledges the support of an Australian Postgraduate Award and a Melbourne University Postgraduate Overseas Research Experience Award.

\label{lastpage}

\end{document}